\documentclass{jpsj-suppl}
\usepackage{txfonts} %Please comment out this line unless the txfonts package is availabe in your LaTeX system.

\title{A Study of the H-dibaryon in Holographic QCD}

\author{Kohei \textsc{Matsumoto}$^{1}$, Yuya \textsc{Nakagawa}$^{1}$, and Hideo \textsc{Suganuma}$^{2}$ }

\inst{$^{1}$Yukawa Institute for Theoretical Physics (YITP), 
Kyoto University, Sakyo, Kyoto 606-8502, Japan \\
$^{2}$Department of Physics, 
Kyoto University, Kitashirakawaoiwake, Sakyo, Kyoto 606-8502, Japan}

\email{kohei.matsumoto@yukawa.kyoto-u.ac.jp}

\recdate{October 2, 2016}

\abst{We study the H-dibaryon (uuddss) in holographic QCD for the first time. Holographic QCD is derived from a QCD-equivalent D-brane system ($S^1$-compactified D4/D8/$\overline{\rm D8}$) in the superstring theory via the gauge/gravity correspondence. In holographic QCD, all baryons appear as topological chiral solitons of Nambu-Goldstone bosons and (axial) vector mesons. In this framework, the H-dibaryon can be described as an SO(3)-type hedgehog state. We present the formalism of the H-dibaryon in holographic QCD, and perform the calculation to investigate its properties in the chiral limit.}

\kword{H-dibaryon, holographic QCD, chiral soliton}

\begin{document}
\maketitle

\section{Introduction}

The H-dibaryon is a $B=2$ SU(3) flavor-singlet bound state of uuddss. In 1977, R.~L.~Jaffe first predicted the existence of H-dibaryons from a group-theoretical argument of the color-magnetic interaction in the MIT bag model \cite{J77}, and estimated the H-dibaryon mass to be $M_{\mathrm{H}} \simeq$ 2150 MeV. 
Since the $\Lambda \Lambda$ threshold is experimentally 2231 MeV, this model calculation implies that the H-dibaryon mass may be stable against the strong decay 
into $\Lambda \Lambda$. 
Several years later, the H-dibaryon was also investigated \cite{BLR85,JK85} in the Skyrme model \cite{S61}, where baryons are described as chiral solitons. This investigation suggests that the H-dibaryon mass is smaller than mass of two nucleons in the chiral limit. This result seems to support Jaffe's prediction. 

However, the prediction of the low-mass H-dibaryon is excluded experimentally in 1991. 
Instead, the double hyper nuclei $_{\Lambda \Lambda}^{~~6}{\rm He}$ was found by Imai group \cite{I91}. In fact, the H-dibaryon is not stable against the strong decay, 
and no H-dibaryon bound state exists at least at the physical point. 
A possible reason of the theoretical failure is due to a large SU(3) flavor-symmetry breaking by the large s-quark mass, $ m_s \gg m_{u,d}$. Then, the current interest is possible existence of the H-dibaryon as a resonance at the physical point. 

Theoretically, there is also an interesting subject of the stability of the H-dibaryon at \textit{unphysical} points such as SU(3) flavor-symmetric cases ($m_u$=$m_d$=$m_s$).
As a resent progress, lattice QCD calculations indicate the existence of the H-dibaryon at some ``unphysical points": the H-dibaryon seems to be stable at the SU(3) flavor-symmetric and large quark-mass region \cite{NPL11,HAL11}. 

Then, how about the H-dibaryon in the chiral limit of $m_u$=$m_d$=$m_s$=0? Although the lattice QCD calculation is usually a powerful tool to evaluate hadron masses, it is difficult to take the chiral limit, since a large-size lattice is required for such calculations. Therefore, for the study of the chiral limit, some model approach \cite{YH16} such as the Skyrme model \cite{S61} would be useful, instead of lattice QCD. Of course, it is desired to use a QCD-based model for the calculation.

In this paper, we study the H-dibaryon and its properties in the chiral limit using holographic QCD \cite{W98,SS05,NSK07}, a recently developed framework to analyze nonperturbative QCD. In particular, we investigate the H-dibaryon mass from the viewpoint of stability and ``existence" of H-dibaryons in the chiral limit.

\section{Holographic QCD}

To begin with, we introduce holographic QCD \cite{W98,SS05}. In the superstring theory, four-dimensional QCD can be constructed using an $S^1$-compactified D4/D8/$\overline{\mathrm{D8}}$-brane system \cite{SS05}, which is called holographic QCD. This QCD-equivalent D-brane system consists of $N_c$ D4-branes and $N_f$ D8/$\overline{\mathrm{D8}}$-branes, which give color and flavor degrees of freedom, respectively. In this construction, gluons and quarks appear as the fluctuation modes of 4-4, 4-8 and 4-$\bar{8}$ strings. This D-brane system has the ${\rm SU}(N_c)$ gauge symmetry and the chiral symmetry \cite{SS05}, and is basically equivalent to QCD in the chiral limit. 

As is often used in holographic QCD, we take $1/N_c$ and $1/\lambda$ expansions, 
where the 't~Hooft coupling $\lambda \equiv N_c g_{\mathrm{YM}}^2$ is expressed with the gauge coupling $g_{\mathrm{YM}}$.
In large $N_c$ and large $\lambda$, $N_c$ D4 branes are replaced by a gravitational background via the gauge/gravity correspondence, and the strong-coupling gauge theory is converted into a weak-coupling gravitational theory \cite{W98}.

In the D4-brane gravitational background, the D8/$\overline{\mathrm{D8}}$ brane system can be expressed with the Dirac-Born-Infeld (DBI) action, 
\begin{equation}
S^{\mathrm{DBI}}_{\mathrm{D8}} = T_8 \int d^9x \:e^{-\phi} \sqrt{-\mathrm{det}(g_{MN}+2\pi \alpha' F_{MN})} \:,
\end{equation}
where $F_{MN}$ is the field strength in the flavor space on the D8 brane, and 
$T_8$, $\phi$ and $\alpha'$ are quantities defined in the superstring theory \cite{SS05}.
From this action, we derive the four-dimensional meson theory equivalent 
to infrared QCD at the leading order of $1/N_c$ and $1/\lambda$ \cite{SS05,NSK07}. 
Here, we only consider massless Nambu-Goldstone (NG) bosons 
and the lightest vector meson ``$\rho$-meson", 
for the construction of low-energy effective theory, 
and finally derive the effective action in 
four-dimensional Euclidean space-time $x^\mu=(t,{\bf x}) $\cite{NSK07}:
\begin{align}
S_{\mathrm{HQCD}} = \int d^4x \:&\Bigl\{
\: \frac{f_{\pi}^2}{4} \mathrm{tr}(L_{\mu}L_{\mu}) 
- \frac{1}{32e^2} \mathrm{tr} [L_{\mu}, L_{\nu}]^2 
+ \frac{1}{2} \mathrm{tr}(\partial_{\mu}\rho_{\nu} - \partial_{\nu}\rho_{\mu} )^2
+ m_{\rho}^2 \mathrm{tr}(\rho_{\mu}\rho_{\mu}) \notag \\
&- i g_{3\rho}\mathrm{tr} \bigl\{ (\partial_{\mu}\rho_{\nu} - \partial_{\nu}\rho_{\mu} )[\rho_{\mu}, \rho_{\nu}] \bigr\} 
- \frac{1}{2}g_{4\rho} \mathrm{tr} [\rho_{\mu}, \rho_{\nu}]^2 
+i g_1 \mathrm{tr} \bigl\{ [\alpha_{\mu}, \alpha_{\nu}] (\partial_{\mu}\rho_{\nu} - \partial_{\nu}\rho_{\mu} ) \bigr\} \notag \\
&+ g_2 \mathrm{tr} \bigl\{ [\alpha_{\mu}, \alpha_{\nu}] [\rho_{\mu}, \rho_{\nu}] \bigr\} 
+ g_3 \mathrm{tr} \bigl\{ [\alpha_{\mu}, \alpha_{\nu}] ([\beta_{\mu}, \rho_{\nu}] +[\rho_{\mu}, \beta_{\nu}] ) \bigr\} \notag \\
&-i g_4 \mathrm{tr} \bigl\{ (\partial_{\mu}\rho_{\nu} - \partial_{\nu}\rho_{\mu} ) ([\beta_{\mu}, \rho_{\nu}] + [\rho_{\mu}, \beta_{\nu}] ) \bigr\} 
- g_5 \mathrm{tr} \bigl\{ [\rho_{\mu}, \rho_{\nu}] ([\beta_{\mu}, \rho_{\nu}] +[\rho_{\mu}, \beta_{\nu}] ) \bigr\} \notag \\
&- \frac{1}{2} g_6 \mathrm{tr} \bigl([\alpha_{\mu}, \rho_{\nu}] + [\rho_{\mu}, \alpha_{\nu}] \bigr)^2 
- \frac{1}{2} g_7 \mathrm{tr} \bigl([\beta_{\mu}, \rho_{\nu}] + [\rho_{\mu}, \beta_{\nu}] \bigr)^2
\: \Bigr\}.
\label{HQCDaction}
\end{align}
Here, $\rho_{\mu}(x) \equiv \rho_{\mu}^a(x)T^a \in {\rm su}(N_f)$ 
denotes the SU($N_f$) lightest vector meson ($\rho$-meson) field, and 
$L_{\mu}$ is defined with the chiral field $U(x)$ as 
\begin{equation}
L_{\mu} \equiv \frac{1}{i}U^{\dagger}\partial_{\mu}U \in {\rm su}(N_f), \quad
U(x) \equiv e^{i2\pi(x)/f_{\pi}} \in {\rm SU}(N_f),
\end{equation}
where $\pi(x) \equiv \pi^a(x)T^a \in {\rm su}(N_f)$ is the NG boson field.
The axial vector current $\alpha_{\mu}$ and 
the vector current $\beta_{\mu}$ are defined as 
\begin{equation}
\alpha_{\mu} \equiv l_{\mu} - r_{\mu},\quad
\beta_{\mu} \equiv \frac{1}{2} ( l_{\mu} + r_{\mu}),
\end{equation}
with the left and the right currents,
\begin{equation}
l_{\mu} \equiv \frac{1}{i}\xi^{\dagger}\partial_{\mu}\xi,\quad 
r_{\mu} \equiv \frac{1}{i}\xi \partial_{\mu}\xi^{\dagger}, \quad
\xi(x) \equiv e^{i\pi(x)/f_\pi}\in {\rm SU}(N_f).
\end{equation}

Thus, we obtain the effective meson theory derived from QCD 
in the chiral limit.
This theory has just two independent parameters, e.g., the Kaluza-Klein mass $M_{\mathrm{KK}} \sim$ 1GeV and $\kappa \equiv \lambda N_c/216\pi^3$ \cite{SS05}, and 
all the coupling constants and masses 
in the effective action (\ref{HQCDaction}) 
are expressed with them \cite{NSK07}. 

Remarkably, in the absence of the $\rho$-meson, 
this effective theory reduces to the Skyrme model \cite{S61} in 
Euclidean space-time:
\begin{equation}
\mathcal{L}_{\mathrm{Skyrme}} = \frac{f_{\pi}^2}{4} \mathrm{tr}(L_{\mu}L_{\mu}) - \frac{1}{32e^2}\mathrm{tr}[L_{\mu},L_{\nu}]^2.
% \: (\: \text{+ WZW term}).
\end{equation}

\section{Topological Chiral Soliton Picture for the H-dibaryon in Holographic QCD}

In general, large-$N_c$ QCD becomes a weakly interacting meson theory, 
and baryons are described as topological chiral solitons of mesons \cite{W79}.
We note that the H-dibaryon is also described as a $B$=2 chiral soliton 
in holographic QCD with large $N_c$, 
like the Skyrme model \cite{BLR85,JK85}.
We study the static H-dibaryon as a $B$=2 chiral soliton 
in holographic QCD, using the ``SO(3)-type hedgehog Ansatz'' 
\begin{equation}
U(\mathbf{x}) = e^{ i\{ ( \mathbf{\Lambda \cdot \hat{x}} )F(r) + 
[( \mathbf{\Lambda \cdot \hat{x}} )^2 - 2/3 ] \varphi(r) \} } 
\in \mathrm{SU(3)}_f, 
\quad F(r) \in {\bf R}, \ \ \varphi(r) \in {\bf R} \quad 
( r \equiv |{\bf x}|, \ \hat{\bf x} \equiv {\bf x}/r) 
\label{SO3HH}
\end{equation}
with the $B$=2 topological boundary condition \cite{BLR85,JK85} of 
\begin{equation}
F(\infty)=\varphi(\infty)=0, \quad F(0)=\varphi(0)=\pi.
\label{BC}
\end{equation}
Here, $ \Lambda_{i=1,2,3}$ are the generators of the SO(3) subalgebra of SU(3)$_f$, 
\begin{equation}
\Lambda_1=\lambda_7=
\begin{pmatrix}
0 & 0 & 0 \\
0 & 0 & -i \\
0 & i & 0 \\
\end{pmatrix},\quad
\Lambda_2=-\lambda_5=
\begin{pmatrix}
0 & 0 & i \\
0 & 0 & 0 \\
-i & 0 & 0 \\
\end{pmatrix},\quad
\Lambda_3=\lambda_2=
\begin{pmatrix}
0 & -i & 0 \\
i & 0 & 0 \\
0 & 0 & 0 \\
\end{pmatrix},
\end{equation}
which satisfy 
%the SO(3) algebraic relation, 
\begin{equation}
[\Lambda_i, \Lambda_j] = i\epsilon_{ijk}\Lambda_k, \quad
\Tr[(\mathbf{\Lambda \cdot \hat{x}})^2-2/3]=0, \quad
(\mathbf{\Lambda \cdot \hat{x}})^3=\mathbf{\Lambda \cdot \hat{x}}.
\end{equation}
Note that $U({\bf x})$ in Eq.(\ref{SO3HH}) 
is the general form of the special unitary matrix 
which consists of $\mathbf{\Lambda \cdot \hat{x}}$.
For the SU(3)$_f$ $\rho$-meson field, 
we use the SO(3) Wu-Yang-'t~Hooft-Polyakov Ansatz, 
similarly in the $B$=1 holographic baryon \cite{NSK07},
\begin{equation}
\rho_0(\mathbf{x})=0,\quad \rho_i(\mathbf{x})
= \epsilon_{ijk} \hat{x_j} G(r) \Lambda_k \in {\rm so}(3) \subset {\rm su}(3),
\quad G(r) \in {\bf R}.
\label{SO3WY}
\end{equation}
In this way, all the above treatments are 
symmetric in the (u, d, s) flavor space.

By substituting Ans\"atze (\ref{SO3HH}) and (\ref{SO3WY}) 
in Eq. (\ref{HQCDaction}), we derive the effective action to describe 
the static H-dibaryon 
in terms of the profile functions $F(r)$, $\varphi(r)$ and $G(r)$ :
\begin{align}
S_{\mathrm{HQCD}} = \int d^4x &\:\biggl\{
\:\: \frac{f_{\pi}^2}{4} \Bigl[\frac{2}{3}\varphi'^2+2F'^2+\frac{8}{r^2}(1 - \cos F \cos \varphi ) \Bigr] \notag \\
&+ \frac{1}{32e^2} \frac{16}{r^2} \Bigl[(\varphi'^2 + F'^2)(1 - \cos F \cos \varphi )
+2\varphi' F' \sin F \sin \varphi \notag \\
&\qquad \qquad \qquad \qquad + \frac{1}{r^2} \bigl\{(1 - \cos F \cos \varphi )^2 + 3\sin^2 F \sin^2 \varphi \bigr\}\Bigr] \notag\\
&+ \frac{1}{2} \Bigl[8 \Bigl( \frac{3}{r^2} G^2 + \frac{2}{r} GG' + G'^2 \Bigr)\Bigr]
+ m_{\rho}^2 [4 G^2] 
+ g_{3\rho}\Bigl[8\frac{G^3}{r}\Bigr] 
+ \frac{1}{2}g_{4\rho}[4G^4] \notag\\ 
&- g_1 \Bigl[ \frac{16}{r} \Bigl\{ \Bigl(\frac{1}{r}G+G' \Bigr) \Bigl(F'\sin \frac{F}{2}\cos \frac{\varphi}{2}+ \varphi' \cos \frac{F}{2} \sin \frac{\varphi}{2} \Bigr) + \frac{1}{r^2}G (1-\cos F \cos \varphi) \Bigr\} \Bigr] \notag\\
&- g_2 \Bigl[\frac{8}{r^2}G^2(1 - \cos F \cos \varphi )\Bigr] \notag\\
&+ g_3 \Bigl[ \frac{16}{r^3} G \Bigl\{ 3\sin F \sin \frac{F}{2} \sin \varphi \sin \frac{\varphi}{2} + \Bigl(1-\cos \frac{F}{2} \cos \frac{\varphi}{2}\Bigr)(1-\cos F \cos \varphi) \Bigr\}\Bigr] \notag \\
&- g_4 \Bigl[ \frac{16}{r^2} G^2 \Bigl(1-\cos \frac{F}{2} \cos \frac{\varphi}{2}\Bigr) \Bigr] 
- g_5 \Bigl[\frac{8}{r}G^3 \Bigl(1-\cos \frac{F}{2} \cos \frac{\varphi}{2}\Bigr) \Bigr] \notag \\
&+ g_6 \bigl[4G^2(F'^2+\varphi'^2)\bigr] 
+g_7 \Bigl[ \frac{8}{r^2} G^2 \Bigl\{ 3\sin^2 \frac{F}{2} \sin^2 \frac{\varphi}{2} + \Bigl(1-\cos \frac{F}{2} \cos \frac{\varphi}{2} \Bigr) ^2 \Bigr\} \Bigr]
\: \biggr\}. 
\label{HQCDFG}
\end{align}

\section{Numerical Results}

Now, to investigate the H-dibaryon in the chiral limit, 
we perform the numerical calculation of the profiles $F(r)$, $\varphi(r)$ and $G(r)$ 
to minimize the Euclidean effective action (\ref{HQCDFG}) 
under the boundary condition (\ref{BC}) \cite{SS98}. 
Here, the two parameters, e.g., $M_{\mathrm{KK}}$ and $\kappa$, 
are set so as to reproduce the pion decay constant $f_{\pi}=92.4 \mathrm{MeV}$ and 
the $\rho$-meson mass $m_{\rho}=776 \mathrm{MeV}$ \cite{SS05, NSK07}.

\begin{figure}[htb]
\vspace{-8 mm}
\begin{tabular}{cc}
\begin{minipage}[t]{0.5\hsize}
\centering
\includegraphics[width=80mm]{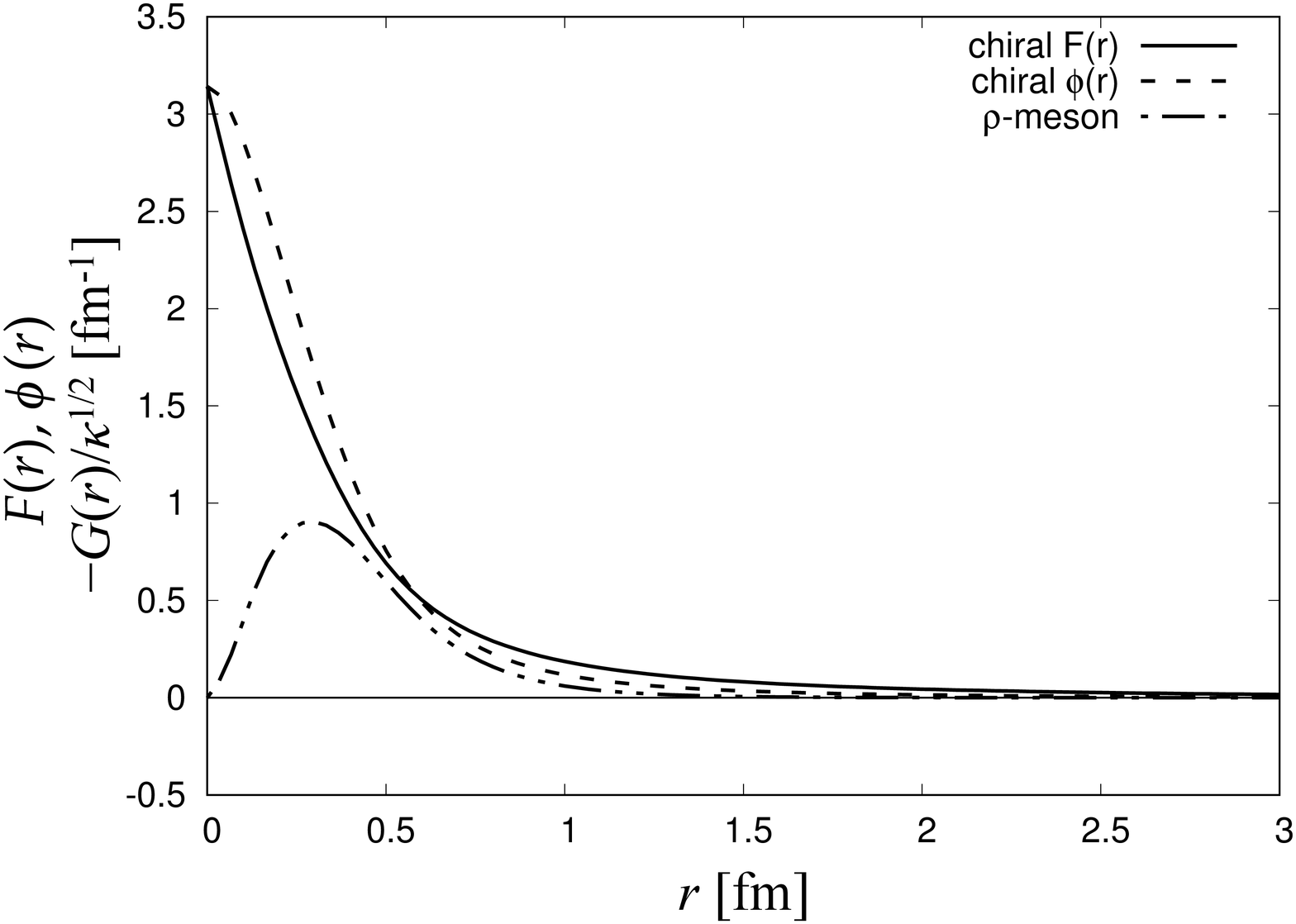}
\caption{The chiral profiles $F(r)$, $\varphi(r)$ and 
the scaled $\rho$-meson profile $G(r)/\kappa^{1/2}$ 
in the H-dibaryon as the SO(3)-type hedgehog soliton 
solution in holographic QCD.}
\label{profile}
\end{minipage} &
\begin{minipage}[t]{0.5\hsize}
\centering
\includegraphics[width=80mm]{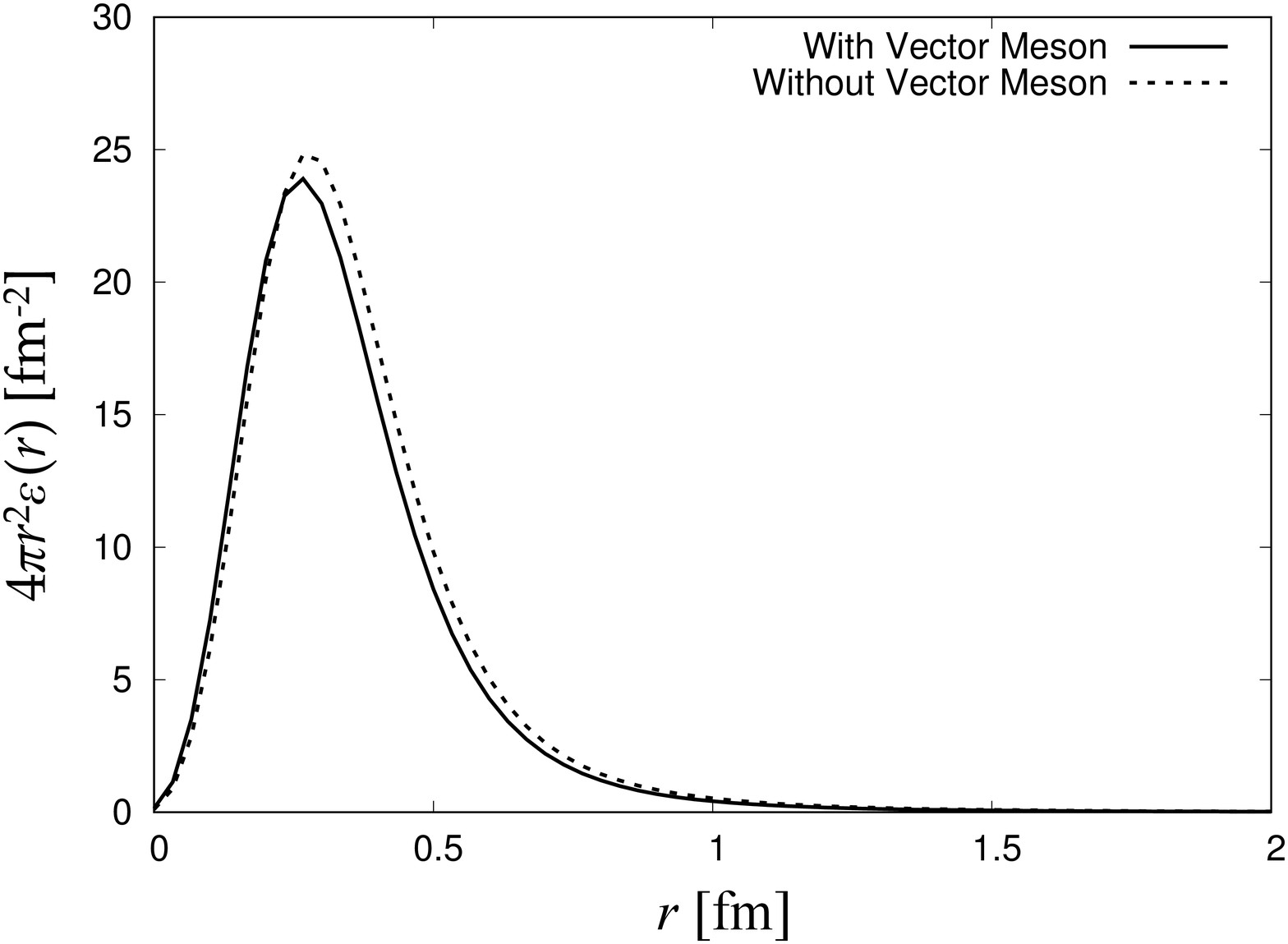}
\caption{The energy density distribution 
$4\pi r^2 \varepsilon(r)$ in the H-dibaryon (solid curve), 
and that without vector mesons (dashed curve) for comparison. }
\label{edvector}
\end{minipage}
\end{tabular}
\vspace{3 mm}
\end{figure}

Figure~\ref{profile} shows the chiral profiles $F(r)$, $\varphi(r)$ and 
the scaled $\rho$-meson profile $G(r)/\kappa^{1/2}$ 
in the soliton solution of the H-dibaryon in holographic QCD.
The H-dibaryon mass is estimated as $M_{\mathrm{H}} \simeq 1673 \mathrm{MeV}$.
We also calculate the energy density 
$4\pi r^2 \varepsilon(r)$ in the H-dibaryon in Fig.~\ref{edvector}, 
and estimate the root mean square radius in terms of the energy density 
as $\sqrt{\langle r^2 \rangle} \simeq$ 0.413fm.
For comparison, we note the $B$=1 hedgehog-baryon mass and radius: 
$M_{B=1}^{\rm HH} \simeq 836.7 \mathrm{MeV}$ and 
$\sqrt{\langle r^2 \rangle} \simeq 0.362 \mathrm{fm}$.
In fact, the H-dibaryon mass is almost equal to two $B$=1 hedgehog-baryon mass,
$M_{\mathrm{H}} \simeq 2.00 M_{B=1}^{\rm HH}$.
Since the nucleon mass $M_{\mathrm{N}}$ is larger than the hedgehog mass $M_{B=1}$ 
by the rotational energy, the H-dibaryon mass is smaller than 
mass of two nucleons (flavor-octet baryons) , $M_{\mathrm{H}} < 2M_{\mathrm{N}}$, 
in the chiral limit.

Finally, we investigate the vector-meson effect for the H-dibaryon.
As the result, we find that the chiral profiles $F(r)$ and $\varphi(r)$ 
are almost unchanged and slightly shrink by the vector-meson effect, 
and the energy density also shrinks slightly, as shown in Fig.~\ref{edvector}.
We find, however, that about 100MeV mass reduction is caused by the vector-meson effect, 
and this mass reduction is due to the interaction between NG bosons and vector mesons 
in the interior region of the H-dibaryon.

\section{Summary}
We have studied the H-dibaryon (uuddss) as the $B$=2 SO(3)-type chiral soliton 
in holographic QCD for the first time.
The H-dibaryon mass is estimated about 1.7GeV in the chiral limit, 
which is smaller than mass of two nucleons (flavor-octet baryons). 
In the H-dibaryon, we have found that, together with slight shrinkage of 
the chiral profile functions $F(r)$, $\varphi(r)$ and the energy density, 
about 100MeV mass reduction of the H-dibaryon is caused by the vector-meson effect.

\section*{Acknowledgements}
We thank Prof. S. Sugimoto and Dr. T. Hyodo for their useful comments and discussions.

\end{document}